\documentclass[aps,prl,a4paper,twocolumn,longbibliography,superscriptaddress
]{revtex4-1}
\usepackage{amsmath}
\usepackage{amssymb}
\usepackage{mathtools}
\usepackage{graphics}
\usepackage{comment}
\usepackage{tikz}

\newcommand{\beq}{\begin{equation}}
\newcommand{\eeq}{\end{equation}}
\newcommand{\bei}{\begin{itemize}}
\newcommand{\eei}{\end{itemize}}
\newcommand{\ben}{\begin{enumerate}}
\newcommand{\een}{\end{enumerate}}

\newcommand{\be}{{\mathbf e}}
\newcommand{\bs}{{\mathbf s}}

\newcommand{\br}{{\mathbf r}}
\newcommand{\bx}{{\mathbf x}}

\newcommand{\vecr}{\mathbf r}

\definecolor{darkblue}{rgb}{0.,0.24,0.51}
\definecolor{britishracinggreen}{rgb}{0.0, 0.26, 0.15}
\definecolor{darkgreen}{rgb}{0,0.60,.2}

\newcommand{\cT}{{\mathcal T}}

\newcommand{\hH}{\hat{H}}
\newcommand{\hO}{\hat{\Omega}}
\newcommand{\hb}{\hat{b}}
\newcommand{\ha}{\hat{a}}

\newcommand{\mO}{\mathcal{O}}

\newcommand{\tphi}{\tilde{\phi}}
\newcommand{\tU}{\tilde{U}}

\newcommand{\ket}[1]{|#1\rangle}
\newcommand{\bra}[1]{\langle #1|}
\newcommand{\braket}[2]{\langle #1 | #2 \rangle}

\usepackage{hyperref}
\hypersetup{
   pdfpagemode=None, 
   pdfstartpage=1,
   pdfmenubar=true,
   pdftoolbar=true,
   colorlinks = true,
   linkcolor=blue,
   citecolor=blue,
   urlcolor=blue,
   bookmarksopen=false
 }

\begin{document}
	\title{Mobile impurity in a Bose-Einstein condensate and the orthogonality catastrophe}
	\author{Nils-Eric Guenther}
	\affiliation{ICFO -- Institut de Ciencies Fotoniques, Barcelona Institute of Science and Technology, 08860 Castelldefels (Barcelona), Spain}
	\author{Richard Schmidt}
	\affiliation{Max-Planck-Institut f\"ur Quantenoptik, Hans-Kopfermann-Str.~1, 85748 Garching, Germany}
	\affiliation{Munich Center for Quantum Science and Technology (MCQST), Schellingstr. 4, 80799 Munich, Germany}
	\author{Georg M.\ Bruun}
		\affiliation{Department of Physics and Astronomy,  Aarhus University, Ny Munkegade, DK-8000 Aarhus C, Denmark}
		\affiliation{Shenzhen Institute for Quantum Science and Engineering and Department of Physics, Southern University of Science and Technology, Shenzhen 518055, China}
	\author{Victor Gurarie}
	\affiliation{Department of Physics, University of Colorado, Boulder CO 80309}
	\author{Pietro Massignan}
		\affiliation{Departament de F\'isica, Universitat Polit\`ecnica de Catalunya, Campus Nord B4-B5, E-08034 Barcelona, Spain}
		\affiliation{ICFO -- Institut de Ciencies Fotoniques, Barcelona Institute of Science and Technology, 08860 Castelldefels (Barcelona), Spain}

	\date{\today}
	
\begin{abstract}
We analyze the properties of an impurity in a dilute Bose-Einstein condensate (BEC). First
the quasiparticle residue of a static impurity in an ideal BEC is shown to vanish with increasing particle number as a stretched exponential, leading to a bosonic orthogonality catastrophe. 
Then we introduce a variational ansatz, which recovers this exact result and describes the macroscopic dressing of the impurity including its back-action onto the BEC as well as boson-boson repulsion beyond the Bogoliubov approximation. 
This ansatz predicts that the orthogonality catastrophe also occurs for mobile impurities, whenever the BEC becomes ideal.  
Finally, we show that our ansatz agrees well with  experimental results. 
\end{abstract}
\maketitle

A single distinguishable particle interacting with a quantum bath, often referred to as the  polaron problem, is one of the simplest realizations of a non-trivial quantum many-body system. 
Its fundamental nature attracted considerable interest since the early days of quantum mechanics, beginning with Landau's seminal paper on electron dressing by phonons \cite{Landau1933}. 
Polaron physics received renewed attention with the advent of ultracold atoms experiments, and it has been extensively studied especially in the case of 
the bath being a Fermi sea, realizing the Fermi polaron. 
Owing to a close interplay between theoretical advances \cite{Chevy2006,Combescot2007,Prokofev2008,Houcke2019} and state-of-the-art experimental 
observations \cite{Schirotzek2009,Kohstall2012,Koschorreck2012,Cetina2016,Scazza2017,Yan2018}, Fermi polarons are now quite well understood even for strong coupling~\cite{Chevy2010,Massignan_Zaccanti_Bruun,FermiPolaronRPP}. 

Recently, the corresponding Bose polaron problem in which impurities are immersed in a dilute BEC has been studied experimentally in three seminal works \cite{Jorgensen2016,Hu2016,Yan2019}. 
Since a bosonic bath is much more compressible than its Fermi analogue and  undergoes a phase
 transition at low temperatures, the problem of an impurity in a BEC exhibits richer few- and many-body physics. Theoretical works so far focused on the coupling of the impurity to Bogoliubov excitations of the BEC \cite{Tempere2009,Rath2013,Christensen2015, Shchadilova2016}, few-body bound states \cite{Yoshida2018,Levinsen2015,Sun2017,Shi2018}, Quantum Monte-Carlo (QMC) studies of the ground state \cite{Ardila2015,Ardila2016,Ardila2018}, as well as finite temperature effects \cite{Levinsen2017,Guenther2018,Dzsotjan2019,Liu2020}. So far, there has, however, been little focus on one fundamental question: what is the fate of the Bose polaron when the interactions in the BEC vanish so that the Bose gas becomes infinitely compressible, allowing for a macroscopic dressing of the impurity? 

In this letter, we carefully analyze this question. First, we show analytically how, for a static impurity in an ideal BEC, the ground state overlap with the non-interacting state vanishes as a stretched exponential with particle number, leading to a bosonic orthogonality catastrophe (OC). 
 We then develop a variational ansatz, which allows  for a macroscopic dressing of the impurity,  
 including the back-action on the BEC as well as  boson-boson repulsion beyond the Bogoliubov approximation. The ansatz, which recovers the exact result for  a static impurity, predicts that the OC also occurs for an impurity  with  \emph{finite} mass when the Bose gas becomes non-interacting. A physical picture emerges where the BEC scatters coherently with the impurity, and a large number of bosons builds a macroscopic but very dilute dressing cloud. Intriguingly, the properties of the polaron are  demonstrated analytically to be given by expressions similar to those obtained from  perturbation theory, even in the regime governed by the OC at  small  boson-boson repulsion. Finally, we show how our ansatz recovers  experimental observations.

{\it A static impurity in an ideal BEC.---}
We start by analyzing a static impurity at zero temperature in an ideal gas of $N$ identical bosons of mass $m_B$ within a sphere of radius $R$. 
For vanishing boson-boson interaction 
the ground state is simply the product $\prod_{j=1}^N\psi(\vecr_j)$ of the lowest energy single-particle wave function. In absence of the impurity, the single particle 
wave function is  $\psi_0(\vecr)=(2\pi R)^{-1/2}\sin(k_0r)/r$
where $k_0= \pi/R$ and $r=|\vecr|$. Introducing an impurity in the center that interacts with the bosons via a potential of the form $U({r})$ with short range $r_{0} \ll R$, the normalized single-particle wave functions for  $r \gg r_{0}$ become
\beq
\label{idealBEC1}
\psi_1(r) =\frac{1}{\sqrt{2 \pi R}\sqrt{1+ \frac{\sin(2\delta)}{2(\pi - \delta)}}
}\frac{\sin \left(k_1 r +\delta \right)}{r}, 
\eeq
where $k_1 = k_0-\delta/R$, and the phase shift $\delta = -\arctan(a k_1)$ is determined by the boson-impurity scattering length  $a$. 
In the thermodynamic limit where $N,R \rightarrow \infty$ at fixed density, both $k_0,k_1 \rightarrow 0$, and for any finite $a<0$ one has $\delta \approx -a k_0 \ll 1$. The difference $\Delta E$ in ground state energies defines the polaron energy~\cite{Massignan2005}
\begin{align}
\Delta E_0 =& N\frac{\hbar^2\left(k_1^2-k_0^2\right)}{2 m_B}= N\frac{\hbar^2 k_0^2}{m_B} \frac{a}{R} = \frac{2 \pi \hbar^2 a}{m_B}n_0 \label{IdealBECDeltaE},
\end{align}
where  $n_0 \equiv N |\psi_0(0)|^2 =  \pi N/2 R^3$ is the BEC density in the center in absence of the impurity.

The overlap between  the ground states with and without the impurity is quantified by the residue $Z_0=|\braket{\Psi_0}{\Psi_1}|^2 = |\braket{\psi_0}{\psi_1}|^{2N}$. Introducing  $k_n=(6 \pi^2 n_0)^{1/3}$, we obtain for large system sizes
\begin{align}
Z_0 =& 
\left[1- \alpha (k_n a)^2/(2N^{2/3})\right]^{2N}
\approx e^{-\alpha N^{1/3}(k_n a)^{2} } \label{IdealBECZ},
\end{align}
where $\alpha \equiv (\pi^2/3+1/4)/(3\pi^3)^{2/3}$. Thus, the residue vanishes as a stretched exponential with increasing particle number, giving rise to a  {\it bosonic orthogonality catastrophe} (bosonic OC). This behavior is even more drastic than the one Anderson predicted for a fermionic bath \cite{AndersonOC}, where the overlap vanishes as a slower power law, due to an infinity of particle-hole excitations in the Fermi sea. The bosonic OC emerges instead because the ideal BEC is infinitely compressible, so that a macroscopic dressing cloud can gather around the impurity.

{\it A mobile impurity in an interacting BEC.--}
We now explore how a finite impurity mass and boson-boson repulsion affect the OC. 
Taking a mobile impurity  of mass $m_I$, the Hamiltonian is
\begin{align} \label{FullHamiltonian}
\hH =\int d^3 r   \left[\hb^\dagger_\br \left(-\frac{\hbar^2\nabla^2_\br}{2 m_B} +\frac{\cT_B}{2}\hb^\dagger_\br \hb_\br-\mu \right) \hb_\br \right.\nonumber \\ 
 \left. + \ha^\dagger_\br\left(-\frac{\hbar^2\nabla^2_\br}{2 m_I}\right)\ha_\br + \int d^3 s \  \hb^\dagger_\br \hb_\br U(\br-\bs)\ha^\dagger_\bs \ha_\bs \right],
\end{align}
where $\ha_\br ^\dagger$ and $\hb_\br^\dagger$ create an impurity and a boson, respectively, at position $\br$, and $\mu$ is the chemical potential. The interaction between  bosons is given by the regularized potential $\cT_B= 4\pi \hbar^2 a_B/m_B$ where $a_B$ is the boson-boson scattering length, which is consistent as long as $n a_B^3 \ll 1$ for any local density $n(\br)$ in the BEC. As before, the interaction between the bath and the impurity is modeled through a 
short-ranged potential $U({\bf r})=U({r})$.
We now  introduce a variational ansatz for the ground state of Eq.~\eqref{FullHamiltonian}, which smoothly connects to the exact result for a static impurity in an ideal BEC. The ansatz includes finite mass effects and
accounts for the boson-boson repulsion beyond the Bogoliubov approximation. To this end, consider the state
\begin{align}
\ket{\Psi} =& \int \frac{d^3 r}{\sqrt{V}} \  \ha^\dagger_\br \exp\left(\int d^3 s \ \phi(\br-\bs)\hb^\dagger_\bs - c.c. \right)\ket{0} \nonumber \\
=& \int \frac{d^3 r}{\sqrt{V}} \ket{\br}\ket{\phi(\br)} \label{CoherentState},
\end{align}
where $V$ is the system volume. It describes a BEC given by the coherent state $\hb_\bs\ket{\phi(\br)} = \phi(\br-\bs)\ket{\phi(\br)}$, which adjusts to the position of the impurity. It follows that $\langle \hat{n}_I(\br)\hat{n}_B(\br+\bs)\rangle\sim|\phi(\bs)|^2$, such that $\phi$ can be regarded as the impurity-bath density-density correlation function. Assuming a  spherically symmetric  ground state $\phi(\br)=\phi(r)$, straightforward algebra yields \cite{SM} 
\begin{align}
&\langle \hH \rangle \! = \!\! \int \! d^3 r \ \phi^*(r)\left[-\frac{\hbar^2\nabla^2_\br}{2 m_r} +U(r)+\frac{\cT_B}{2}|\phi(r)|^2-\mu \right]\phi(r) \label{EFunctional}
\end{align}
with $m_r^{-1}=m_B^{-1}+m_I^{-1}$ the reduced mass. 
Minimizing Eq.~\eqref{EFunctional} gives
\begin{align}\label{extendedGPE}
 \left[-\frac{\hbar^2\nabla^2_\br}{2 m_r}+U(r)+\cT_B |\phi(r)|^2\right] \phi(r) = \mu \phi(r).
\end{align} 
 Equation \eqref{extendedGPE} is remarkably simple: The mobile impurity is described by a modified Gross-Pitaevskii equation (GPE) where the kinetic term contains the reduced mass, and the bosons scatter collectively on the impurity.  
 Note that the back-action on the BEC due to the dressing cloud around the impurity and boson-boson interactions are naturally included in the GPE, an effect that is not fully accounted for in a Bogoliubov approach to the problem that expands in fluctuations around a homogenous BEC \cite{Shchadilova2016,Dzsotjan2019}.  
 For infinite impurity mass we have $m_r=m_B$, and Eq.~\eqref{extendedGPE} reduces to the standard GPE for a static potential $U(r)$. If, moreover, $a_B=0$ it reduces to the one-body Schr\"odinger equation thereby recovering  the exact results for  the bosonic OC described above.

To investigate the ground state of the impurity, we solve Eq.~\eqref{extendedGPE} subject to the condition $\mu = n_0 \cT_B$, which ensures that the density far away from the impurity converges to the density $n_0$ of the BEC in absence of the impurity.
Introducing the dimensionless quantities $\bx = \br/\xi$, $\tphi(x)= \phi(x \xi)/\sqrt{n_0}$ and $\tU(x) = 2 m_r \xi^2 U(x \xi)/\hbar^2$ with $\xi = (8\pi n_0 a_B m_r/m_B)^{-1/2}$, Eq.~\eqref{extendedGPE} becomes 
\begin{align}\label{GPE_adim}
\left[-\nabla^2_x+\tU(x)+|\tphi(x)|^2-1\right]\tphi(x) = 0,
\end{align}
showing that the generalized healing length $\xi$ is the natural length scale of the problem.  

The polaron energy is the energy shift $\Delta E$ away from the solution with $\tU=0$. Using Eq.~\eqref{GPE_adim} in \eqref{EFunctional}, one finds 
\begin{align}\label{DeltaE_adim}
\Delta E = -\frac{E_\xi}{2} \int d^3 x \ (|\tphi(x)|^4-1),
\end{align}
where $E_\xi=\hbar^2 n_0 \xi/2 m_r $. 
The polaron quasiparticle weight $Z$ is the overlap between ground states of the gas with and without impurity. With the coherent ansatz one finds
\begin{align}\label{Z_adim}
Z=|\braket{\Psi_0}{\Psi_1}|^2 = e^{-N_\xi \int d^3 x \ |\tphi(x)-1|^2 },
\end{align}
where we defined $N_\xi= n_0 \xi^3$. This shows that changes in the condensate mode cause an exponential suppression of the overlap. 
Finally, the number $\Delta N$ of 
 bosons in the dressing cloud around the impurity is
\begin{align}\label{DeltaN_adim}
\Delta N = N_\xi \int d^3 x \ (|\tphi(x)|^2-1).
\end{align}

{\it Condensate wave function.---}
The condensate wave function $\phi(\bx)$ is obtained numerically. In our computation we employ three attractive potentials: 
a square well $U_w \Theta(1-r/r_0)$, a Gaussian $U_g \exp(-r/r_0)^2$, and an exponential  $U_e \exp(-r/r_0)$, which give rise to effective ranges $r_e\sim r_0$, mimicking open-channel dominated resonances \cite{Chin2010}. We tune
 their depth $U$ and characteristic size $r_0$ independently, so that we can model different scattering lengths $a$ at fixed effective range $r_e$. Within the wide range of parameters explored, the results given by these three potentials differ by less than the width of the lines in all figures, demonstrating an effective two-parameter universality (given by the scattering length $a$ and range $r_e$) governing this problem. 
 Note that in general one cannot use a zero-range potential for both the boson-boson and the boson-impurity interaction,  
because in that case Eq.~\eqref{extendedGPE} admits only a zero-energy polaron solution.

Numerically, the problem needs careful treatment, because next to the impurity the wave function varies  on scales comparable to $r_e$, while further away it evolves on a scale set by $\xi\gg r_e$. To achieve sufficient accuracy despite this large separation of scales, we discretize the integral in Eq.~\eqref{EFunctional} on a  non-uniform grid featuring an exponentially-growing lattice spacing in the outward radial direction containing several thousand points both inside and outside the potential.
For all computational results presented here, we used a grid with maximal radius $R= 100\xi$, and boundary condition $\phi(R)=\sqrt{n_0}$. 

\begin{figure}[t]
	\includegraphics[width=\columnwidth]{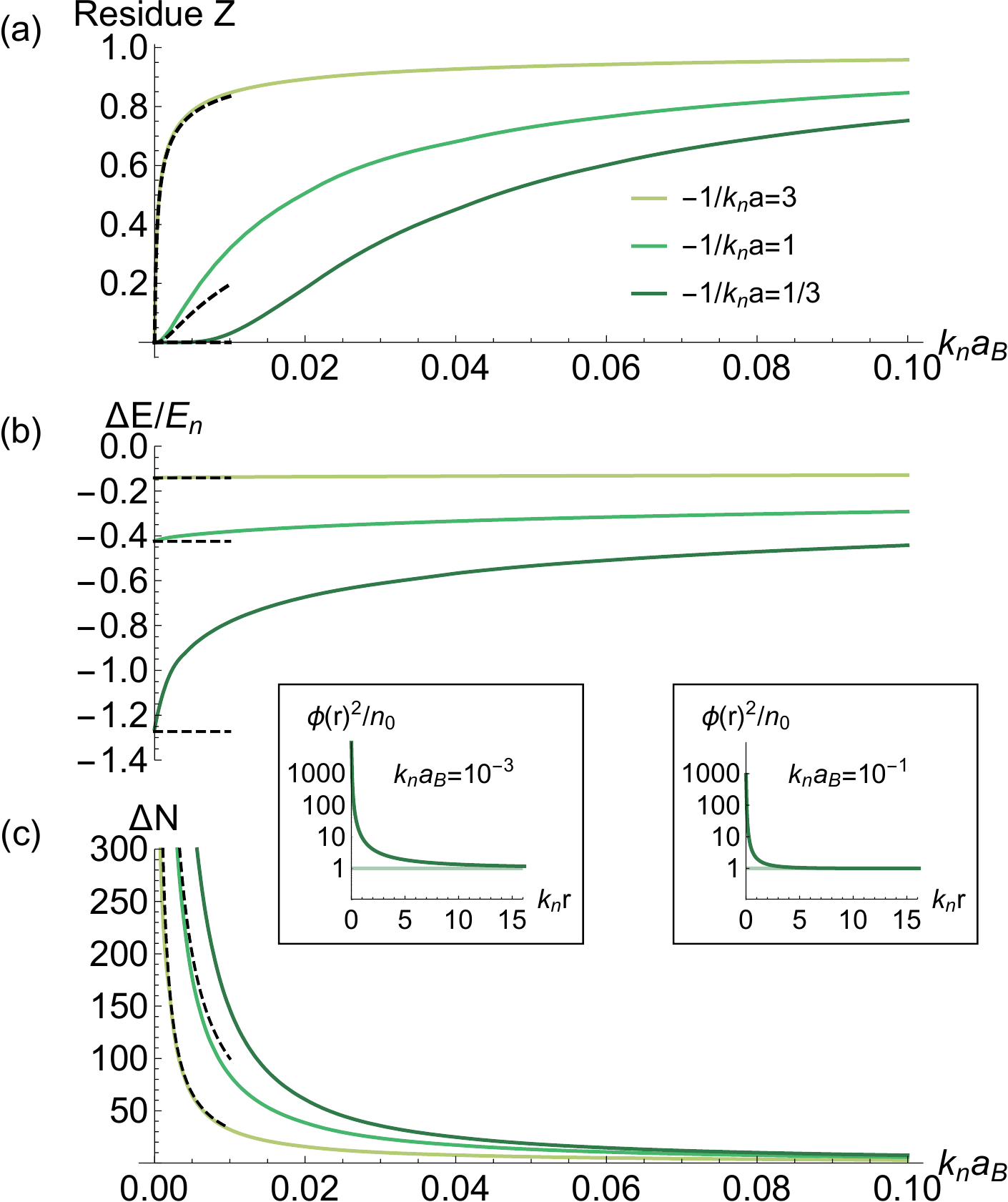}
	\caption{\label{fig:Orthogonality_2}
\textbf{Polaron properties.} (a) The residue and (b) the energy (in units $E_n = \hbar^2 k_n^2/2m_B$) 
of the polaron, as well as (c) the number of particles in the dressing cloud as a function of $k_n a_B$ for $-1/k_n a = (3, 1, 1/3)$ (light green to darker green lines). In all panels, $m_I=m_B$ and $k_n r_e=0.05$.
The dashed lines are the perturbative expressions Eqs.~\eqref{DeltaEpert}-\eqref{DeltaNpert}. 
 The insets show $|\phi(r)|^2$ for $-1/k_na=1/3$ and for two values of $k_na_B$, corresponding to $k_n \xi \approx {70}$ (left) and $k_n \xi \approx 7$ (right).
}
\end{figure}

{\it Results.---}
In Fig.~\ref{fig:Orthogonality_2} we plot the residue and energy of the polaron as well as the number of particles in its dressing cloud as a function of the boson-boson scattering length $a_B$ for various impurity-boson interaction strengths. 
Fig.~\ref{fig:Orthogonality_2}(a)  shows that, in contrast to fermions, the OC persists even for mobile impurities when the BEC becomes ideal. A related finding was discussed in Refs.~\cite{Shchadilova2016,Yoshida2018}. 
In this limit, the residue $Z$ vanishes and the number of particles $\Delta N$ in the dressing cloud diverges for $k_na_B\rightarrow 0_+$, see Fig.~\ref{fig:Orthogonality_2}($b$). The bosonic OC is cured when the bosons start to repel which leads to a suppression of particles in the dressing cloud.

Equation~\eqref{GPE_adim} can be solved analytically for  $|a|^3\ll \xi^2 r_0$~\footnote{see N.-E. Guenther {\it et al.}, in preparation}. Under this condition, which is  fulfilled for any finite $|a|$ when $a_B\rightarrow 0$, one obtains for $x \gtrsim r_0/\xi$ 
 the Yukawa solution  
 \begin{align}
\tphi(x) = 1- (a/\xi)e^{-\sqrt{2}x}/x. \label{GPE_Yukawa}
\end{align}
Using this in Eqs.~\eqref{DeltaE_adim}-\eqref{DeltaN_adim} gives 
\begin{alignat}{2}
\Delta E =& \ 4\pi  E_\xi a/\xi &&=2\pi \hbar^2 a n_0/m_r,  \label{DeltaEpert} \\
\log Z =& -\sqrt{2}\pi  N_\xi a^2/\xi^2&&=- \sqrt{2}\pi n_0 \xi a^2, \label{Zpert}\\
\Delta N =& -4\pi  N_\xi a/\xi&&=-a m_B/2 a_B m_r. \label{DeltaNpert}
\end{alignat}
These  expressions, which are recovered by our numerical results when $k_na_B\rightarrow 0$ (so that $a/\xi \to 0$) as shown in Fig.~\ref{fig:Orthogonality_2}, analytically describe how the Bose polaron disappears  in the limit of an ideal Bose gas.  In particular, the residue vanishes exponentially with  $\log Z\propto -1/\sqrt{a_B}$ 
and the number of particles in the polaron cloud grows as $\Delta N\propto 1/a_B$ when the BEC looses its stiffness, leading to the build-up of a macroscopic screening cloud around the impurity and causing the bosonic OC.
Intriguingly, Eqs. \eqref{DeltaEpert} and \eqref{Zpert} have the same functional form as those obtained from an expansion in $k_n a$ \cite{Christensen2015,Casteels2014}, even though they are valid close to the bosonic OC, which must be expected to be well beyond the radius of convergence of perturbation theory.
 
The insets in  Fig.~\ref{fig:Orthogonality_2} display $|\phi(r)|^2$ for $1/k_na =-1/3$ at  boson-boson scattering lengths $k_na_B=0.001$ and $0.1$. They show how the bosons pile up around the impurity in a macroscopic dressing cloud of size $\sim\xi$. 
Importantly, in this case the local gas parameter $|\phi(r)|^2 a_B^3$  remains  small \textit{everywhere}, even close to the impurity.
In the most strongly interacting case, the unitary limit $|k_n a|\rightarrow \infty$, we found it to vanish as $\propto k_n a_B (a_B/r_e)^{4/3}$ when $k_na_B\rightarrow 0$. This ensures that the assumption of a contact potential for the boson-boson interaction in Eq.~\eqref{FullHamiltonian} is consistent. It also shows that the large number of bosons in the dressing cloud is due to a large radius $\sim \xi$ and not to an exceedingly large density.  

{\it Comparison with experiment.--} We now compare the predictions of our ansatz with  experiments. Close to a Feshbach resonance, the effective range  $r_e$, defined through the low-energy expansion of the phase shift 
$k\cot \delta = -1/a+r_e k^2/2+O(k^2)$, varies slowly around its value right at resonance~\cite{Chin2010,Schmidt2012,Viel2016,Tanzi2018}
\beq
r_e=-2R^*+2\Gamma(1/4)^2R_{\rm vdW}/(3\pi),
\eeq
Here, $R^*=\hbar^2/2m_r a_{\rm bg}\delta\mu\,\Delta B$, $\Gamma(x)$ is the Gamma function, 
and $R_{\rm vdW}$ is the van der Waals radius. For open-channel dominated resonances, one has $R_{vdW}\gg R^*$ such that $r_e \sim R_{vdW} \ll \xi$ in typical experiments. For example, 
the experiments in Aarhus \cite{Jorgensen2016}, JILA \cite{Hu2016}, and MIT \cite{Yan2019}, featured $k_n \xi= 21.7,\ 8,\ 9.6$, respectively, giving $r_e/\xi=0.002, \ 0.02, \ 0.01$ (using data from Refs.~\cite{Tanzi2018,Viel2016,Zirbel2008}). 
 
\begin{figure}[t!]
	\begin{center}
		\includegraphics[width=\columnwidth]{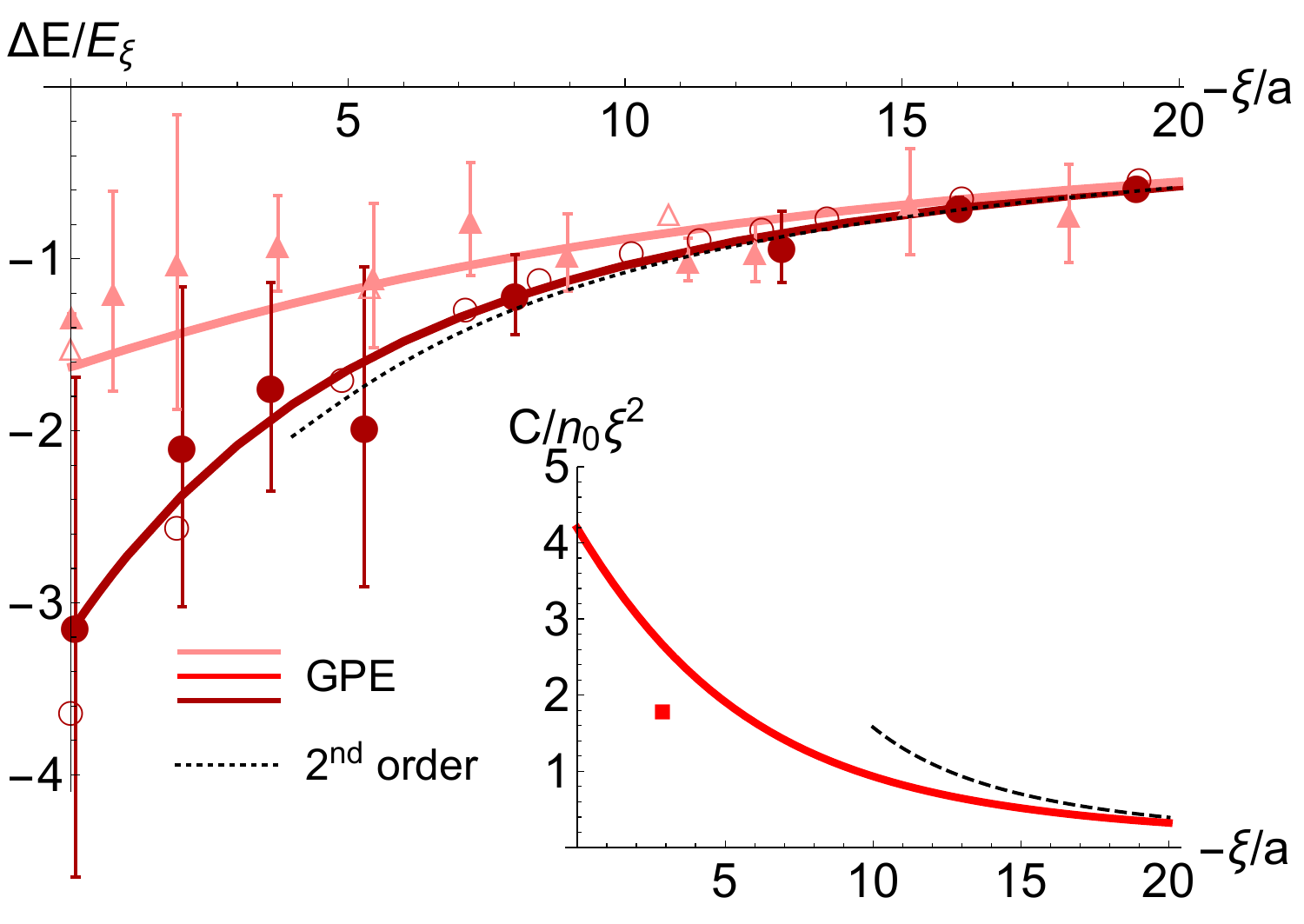}
	\end{center}
	\caption{\label{fig:energyAcrossResonance}
		\textbf{Polaron energy and contact.} Polaron energy $\Delta E$ in units of $E_\xi = \hbar^2n_0\xi/2m_r$ as a function of $-\xi/a$. The solid lines are the results of our ansatz
		obtained for effective ranges $r_e/\xi=0.002$ (pink) and $0.02$ (dark red), corresponding respectively to the experimental conditions of Aarhus \cite{Jorgensen2016} and JILA \cite{Hu2016}.
		Filled symbols are the experimental data from Aarhus (triangles) and JILA (circles), and empty symbols are the corresponding QMC data from \cite{Ardila2016,Ardila2018}. The black dotted line is the second order perturbative result $\Delta E = 4\pi(a/\xi)(1+\sqrt{2}a/\xi)E_\xi$~\cite{Casteels2014,Christensen2015}. The inset shows Tan's contact $C$ of a single impurity given by Eq.~\eqref{Contact}, as a function of $-\xi/a$ for the effective range $r_e/\xi=0.01$. The red point is the measurement reported by the MIT group~\cite{Yan2019} at lowest temperature and closest to unitarity, using a resonance with a comparable ratio $r_e/\xi$. The dashed line is the perturbative result, Eq.~\eqref{Cpert}.
	}
\end{figure}

 Our numerical results for $\Delta E$ are shown in Fig.~\ref{fig:energyAcrossResonance}
 together with the measurements from the experiments reported in Refs.~\cite{Jorgensen2016,Hu2016}, which had mass ratios 
 $m_I/m_B=1 $ and $40/87$, respectively. Corresponding QMC results of Refs.~\cite{Ardila2016,Ardila2018} are also displayed. 
The coherent state ansatz shows good agreement with the experimental  data, even for impurities with mass comparable to $m_B$ or lighter.
Our ansatz predicts large dressing clouds at unitarity, containing $20-180$ bosons for the JILA and Aarhus case, respectively. This  strong dressing, however, is accompanied by an extremely small residue $Z$ at resonance and therefore a very small spectral weight of the ground state in  the experimental radio-frequency spectrum at odds with other theories that recover the experimentally measured spectrum~\cite{Rath2013,Levinsen2015}.

{\it Contact.--} We finally examine Tan's contact parameter, which quantifies the short-range correlations between the impurity and the atoms
in the BEC. It can be obtained from Tan's adiabatic theorem~\cite{Tan2008a,Tan2008b,Braaten2008,Braaten2010,Werner2012}
\begin{align}\label{Contact}
C= -\frac{8 \pi m_r}{\hbar^2}\frac{\partial(\Delta E)}{\partial (1/a)}
 = -4\pi n_0\xi^2 \frac{\partial (\Delta E/E_\xi)}{\partial (\xi/a)}.
\end{align}
Our ansatz gives to leading order in $a/\xi$
\begin{align}
C_1= (4 \pi a/\xi)^2 n_0\xi^2=16 \pi^2 n_0 a^2, \label{Cpert}
\end{align}
which agrees with the leading order result of perturbation theory. 
In the inset of Fig.~\ref{fig:energyAcrossResonance} we show the dependence of the contact as function of $-\xi/a$ for ratios $r_e/\xi$ and mass ratios appropriate for the experiment at MIT \cite{Yan2019}. In this experiment the contact, shown as a red square, was obtained from the tail of the radio-frequency response at finite temperatures.

{\it Discussion and outlook.--}
Using an ansatz describing the  macroscopic dressing of the impurity and the back-action on the BEC including the boson-boson repulsion beyond the Bogoliubov approximation, we carefully analyzed the fate of the polaron with decreasing boson-boson interaction. We showed  that the polaron disappears for $a_B\rightarrow 0$ resulting in  a bosonic orthogonality catastrophe also when it has a finite mass. 
Strikingly, our ansatz predicts that the properties of the polaron are accurately described by expressions similar to  perturbation theory even in a regime where the polaron picture ceases to be valid, and perturbation theory becomes formally invalid. It would be very interesting to examine this experimentally for instance  using a Feshbach resonance to tune $a_B$, and 
employing Rabi \cite{Kohstall2012} or Ramsey \cite{Cetina2015,Cetina2016,Ashida2018} spectroscopy. 
Also, the predicted large dressing clouds suggest potentially strong induced impurity-impurity interactions, which could affect the spectrum 
even for small impurity concentrations~\cite{Camacho_Guardian2018,Tan2020}.

{\it Note added.---} During submission of this manuscript, we learned of interesting parallel theoretical work focusing on the dynamical properties of an impurity, which derived a similar expression for the ground state energy using the Lee-Low-Pines  transformation, and introduced a finite range in the bose-bose interaction \cite{Drescher2020}.

\vspace{5mm}
\begin{acknowledgments}
We acknowledge inspiring and insightful discussion with G. Astrakharchik, T. Enss, J. Levinsen, M. Lewenstein, M. Parish and L. Tarruell, and we thank N. B. J\o rgensen for providing experimental data from Ref.~\cite{Jorgensen2016}.
N.G. is supported by a ``la Caixa-Severo Ochoa'' PhD fellowship. 
N.G. acknowledges the Spanish Ministry MINECO (National Plan 15 Grant: FISICATEAMO FIS2016-79508-P, SEVERO OCHOA SEV-2015-0522, FPI), European Social Fund, Fundaci\'o Cellex, Fundaci\'o Mir-Puig, Generalitat de Catalunya (AGAUR Grant 2017 SGR 1341 and CERCA/Program), ERC AdG NOQIA, EU FEDER, MINECO-EU QUANTERA MAQS, and the National Science Centre Poland (Symfonia Grant 2016/20/W/ST4/00314).
R.S. is supported by the Deutsche Forschungsgemeinschaft (DFG, German Research Foundation) under Germany's Excellence Strategy -- EXC-2111 -- 390814868.
G.M.B.\ acknowledges financial support from the Independent Research Fund Denmark - Natural Sciences via Grant No. DFF - 8021- 00233B. 
This work was supported by the Simons Collaboration on Ultra-Quantum Matter, which is a grant from the Simons Foundation (651440, VG).
P.M. acknowledges the ``Ram\'on y Cajal" program, Spanish MINECO (FIS2017-84114-C2-1-P), and EU FEDER Quantumcat.
\end{acknowledgments}

\bibliography{CoherentImpurity}

\onecolumngrid
\begin{center}
\newpage
\textbf{
Supplemental Material:\\[4mm]
\large Mobile impurity in a Bose-Einstein condensate and the orthogonality catastrophe}\\
\vspace{4mm}
{Nils-Eric Guenther,$^1$ Richard Schmidt,$^{2,3}$ Georg M.\ Bruun,$^{4,5}$ Victor Gurarie,$^6$ and Pietro Massignan,$^{7,1}$}\\
\vspace{2mm}
{\em \small
$^1$ICFO -- Institut de Ciencies Fotoniques, Barcelona Institute of Science and Technology, 08860 Castelldefels (Barcelona), Spain\\
$^2$Max-Planck-Institut f\"ur Quantenoptik, Hans-Kopfermann-Str.~1, 85748 Garching, Germany\\
$^3$Munich Center for Quantum Science and Technology (MCQST), Schellingstr. 4, 80799 Munich, Germany
$^4$Department of Physics and Astronomy,  Aarhus University, Ny Munkegade, DK-8000 Aarhus C, Denmark\\
$^5$Shenzhen Institute for Quantum Science and Engineering and Department of Physics, Southern University of Science and Technology, Shenzhen 518055, China\\
$^6$Department of Physics, University of Colorado, Boulder CO 80309\\
$^7$Departament de F\'isica, Universitat Polit\`ecnica de Catalunya, Campus Nord B4-B5, E-08034 Barcelona, Spain
}
\end{center}

\setcounter{equation}{0}
\setcounter{figure}{0}
\setcounter{table}{0}
\setcounter{section}{0}
\setcounter{page}{1}
\makeatletter
\renewcommand{\theequation}{S.\arabic{equation}}
\renewcommand{\thefigure}{S\arabic{figure}}
\renewcommand{\thetable}{S\arabic{table}}
\renewcommand{\thesection}{S.\arabic{section}}

\section*{Derivation of the modified GP equation}\label{app:EnergyFunctional}
We start by presenting a step-by-step derivation of the expectation value of the grand potential,
\beq\label{FullHamiltonianSM}
\hO =\int d^3 r  \left[\hb^\dagger_\br \left(-\frac{\hbar^2\nabla^2_\br}{2 m_B} +\frac{\cT_B}{2}\hb^\dagger_\br \hb_\br-\mu \right) \hb_\br + \ha^\dagger_\br\left(-\frac{\hbar^2\nabla^2_\br}{2 m_I}\right)\ha_\br + \hb^\dagger_\br \hb_\br U(\br-\bs)\ha^\dagger_\bs \ha_\bs \right],
\eeq
upon the ansatz state
\beq \label{CoherentStateSM}
\ket{\Psi} = \int \frac{d^3 r}{\sqrt{V}} \  \ha^\dagger_\br \exp\left(\int d^3 s \ \phi(\br-\bs)\hb^\dagger_\bs - c.c. \right)\ket{0}
= \int \frac{d^3 r}{\sqrt{V}} \ket{\br}\ket{\phi(\br)}.
\eeq
The result contains three contributions, coming from the bath, the impurity, and their mutual interaction.
The contribution from the bath is very simple. 
Using $\braket{\br}{\br'}=\delta(\br-\br')$ and $\hb_\bs \ket{\phi(\br)} = \phi(\br-\bs)\ket{\phi(\br)}$, one finds:
\begin{align}
\langle \hO_B \rangle =& \int \frac{ d^3 r \ d^3 s}{V} \phi^*(\br-\bs)\left[-\frac{\hbar^2 \nabla^2_\bs}{2 m_B} + \frac{\cT_B}{2}|\phi(\br-\bs)|^2-\mu\right] \phi(\br-\bs)
= \int d^3 r \ \phi^*(\br)\left[ -\frac{\hbar^2\nabla^2_\br}{2 m_B} +\frac{\cT_B}{2} |\phi(\br)|^2 -\mu \right]\phi(\br).
\end{align}
Similarly, the bath-impurity interaction term gives:
\begin{align}
\langle \hO_\text{int}\rangle =& \int \frac{d^3 r \ d^3 s }{V} \ U(\br-\bs) \bra{\phi(\br)}\hb_\bs^\dagger \hb_\bs \ket{\phi(\br)}
= \int \frac{d^3 r \ d^3 s}{V} U(\br-\bs) |\phi(\br-\bs)|^2
=\int d^3 r \ U(\br) |\phi(\br)|^2.
\end{align}
To study the impurity sector, care must be taken when evaluating the Laplacian. A possible approach is working explicitly with a difference quotient
\begin{align}
\nabla^2_\br \ha_\br = \lim_{d\rightarrow 0} \sum_{i=(1,2,3)} \frac{\ha_{\br+ d \be_i}+ \ha_{\br-d\be_i}-2 \ha_\br}{d^2}, \label{OLap}
\end{align}
and taking the limit $d\rightarrow0$ at the end. Here $\be_i$ is any orthonormal basis set. 
This gives
\begin{align}
\langle\hO_I\rangle =& -\frac{\hbar^2}{2 m_I} \int d^3 r \lim_{d\rightarrow 0} \sum_{i=(1,2,3)}  \langle \ha^\dagger_\br \frac{\ha_{\br+ d \be_i}+ \ha_{\br-d\be_i}-2 \ha_\br}{d^2}\rangle \nonumber \\
=& - \frac{\hbar^2}{2 m_I} \lim_{d\rightarrow 0} \int \frac{d^3 r}{V d^2} \sum_{i=(1,2,3)} \left( \braket{\phi(\br)}{\phi(\br+ d \be_i)}+\braket{\phi(\br)}{\phi(\br-d \be_i)}-2\underbrace{\braket{\phi(\br)}{\phi(\br)}}_{1} \right). \label{KineticImpurity}
\end{align}
The overlap between different coherent states is
\begin{align}\label{Overlap_Coherent}
\braket{\phi_1}{\phi_2}= \exp \left[ \int d^3 r \left( \ \phi_1^*(\br)\phi_2(\br)- \frac{|\phi_1(\br)|^2}{2}-\frac{|\phi_2(\br)|^2}{2} \right)\right].
\end{align}
Applied to the last line in Eq.~\eqref{KineticImpurity}, this gives
\begin{align}
\braket{\phi(\br)}{\phi(\br\pm d \be_i)} =& \exp \left[ \int d^3 s \left(\phi^*(\br-\bs)\phi(\br \pm d\be_i-\bs)-\frac{|\phi(\br-\bs)|^2}{2}-\frac{|\phi(\br\pm d\be_i-\bs)|^2}{2}   \right) \right]\nonumber \\
=& \exp \left[\int d^3 s \left( \phi^*(\bs)\phi(\bs\mp d \be_i)-|\phi(\bs)|^2 \right) \right]\\
=& 1 \mp d \underbrace{\int d^3 s \ \phi^*(\bs) \partial_i \phi(\bs)}_{=0} + \frac{d^2}{2} \left[\int d^3 s \ \phi^*(\bs)\partial_i^2 \phi(\bs) + \underbrace{\left(\int d^3 s \ \phi^*(\bs)\partial_i\phi(\bs) \right)^2}_{=0} \right] + \mO(d^3) \label{Overlapdh},
\end{align}
where the first and last term vanish because $\phi^*(\bs)\partial_{i}\phi(\bs)$ is antisymmetric in $s_i$ due to $\phi$ being a central function, $\phi(\bs)= \phi(s)$. Plugging this expression into \eqref{KineticImpurity} gives:
\begin{align}
\langle \hat{\Omega}_I\rangle = \int d^3 r \ \phi^*(\br)\left( -\frac{\hbar^2\nabla^2_\br}{2 m_I} \right)\phi(\br) \label{KineticImpurityFinal}.
\end{align}
Combining all terms, the expectation value of $\hO$ over the coherent ansatz $\ket{\Psi}$ reads:
\begin{align}
\langle \hO\rangle = \int d^3 r \ \phi^*(\br)\left(-\frac{\hbar^2\nabla^2_\br}{2 m_r} +U(\br)+ \frac{\cT_B}{2}|\phi(\br)|^2-\mu \right)\phi(\br)\label{relativeGPEnergyAppendix},
\end{align}
where $m_r^{-1}=m_B^{-1}+m_I^{-1}$ is the reduced mass for one bath boson and one impurity. By minimizing \eqref{relativeGPEnergyAppendix} with respect to $\phi^*$, one recovers the equation introduced in the text,
\begin{align}
 \left[-\frac{\hbar^2\nabla^2_\br}{2 m_r}+U(r)+\cT_B |\phi(\br)|^2\right] \phi(\br) = \mu \phi(\br),
\end{align}
which is a modified GP equation (due to the presence of the reduced mass $m_r$) describing the coherent dressing of a mobile impurity when it is immersed in a weakly-interacting BEC.

\end{document}